\documentclass{astlb}

\usepackage[T2A]{fontenc}
\usepackage[utf8]{inputenc}

\usepackage{lscape}
\usepackage{xspace} 
\usepackage{url}

\usepackage{graphicx}	
\usepackage{amssymb}

\def\lum{erg\,s$^{-1}$}
\def\integral{{INTEGRAL}}
\def\ibis{{IBIS}}

\def\flux{erg s$^{-1}$ cm$^{-2}$}

\begin{document}

\journalinfo{(2024)}{50}{1}{25}[33]

\title{Deep Hard X-ray Survey of the M81 Field Based on INTEGRAL Data}

\author{
R.~A.~Krivonos\address{1}\email{krivonos@cosmos.ru},
I.~A.~Mereminskiy\address{1},
S.~Yu.~Sazonov\address{1},
\addresstext{1}{\it Space Research Institute, Russian Academy of Sciences, Moscow, 117997 Russia}
}

\shortauthor{Krivonos et al.}

\shorttitle{INTEGRAL/IBIS hard X-ray survey of M81 field}

\submitted{21.11.2023 \\
Revised November 21, 2023; Accepted November 21, 2023}

\begin{abstract}
We have carried out a deep survey of the M81 field in the 25$-$60 keV energy band based on long-term (2003$-$2023) INTEGRAL observations. A record sensitivity of 0.16 mCrab at a detection significance of $4\sigma$ has been achieved in the central part of the field owing to the long accumulated exposure (19.2 Ms). The total area of the survey is 1004 deg$^2$ at a sensitivity level better than 0.72 mCrab. We have produced a catalog of sources detected at a significance level higher than $4\sigma$. It contains 51 objects most of which are active galactic nuclei (AGNs). The median redshift of the Seyfert galaxies in the catalog is $z=0.0366$. Six sources have not been detected previously in any of the X-ray surveys. According to the available indirect data, all of them and two more sources that have already been entered previously into the INTEGRAL survey catalogs can also be AGNs, including those with strong internal absorption.


\end{abstract}

\section*{INTRODUCTION}

The International Gamma-Ray Astrophysics Laboratory \citep[\integral,][]{2003A&A...411L...1W} is a project of the European Space Agency in collaboration with Roskosmos and NASA. The observatory was launched on October 17, 2002, from the Baikonur Cosmodrome by a PROTON rocket. A combination of the high sensitivity in the energy range  ${\sim}20-100$~keV, the large field of view ($28^{\circ}\times28^{\circ}$), and the relatively good angular resolution ($12'$) of the \ibis\ coded-mask telescope \citep{2003A&A...411L.131U} gives great opportunities to investigate the sky in hard X-rays. Over more than 20 years of its in-orbit operation the INTEGRAL observatory has carried out a lot of observations in many different parts of the sky \citep{2022MNRAS.510.4796K}.

The X-ray surveys conducted with the \integral\ telescopes serve as a basis for systematic studies of various classes of objects: cataclysmic variables and symbiotic stars \citep{2020NewAR..9101547L}, low-mass and high-mass X-ray binaries \citep{2020NewAR..8801536S,2019NewAR..8601546K}, and active galactic nuclei \citep[AGNs,][]{2020NewAR..9001545M}.

In this paper we present the results of  deep survey of the sky region around the nearby M81 group of galaxies \citep{2002A&A...383..125K} constructed from the INTEGRAL data accumulated in various observations of this field in the entire time of its in-orbit operation. The central galaxy of the group, M81, is located at a distance of 3.7 Mpc \citep{2018MNRAS.479.4136K}.

\section{THE M81 FIELD}
\label{sec:m81}

For the \integral\ observatory the extragalactic M81 ﬁeld is unique in accumulated exposure and achieved sensitivity. Initially, this ﬁeld was observed as part of the programs devoted to investigating the spectra of ultraluminous X-ray sources in nearby galaxies \citep[M82 X-1, Hol IX X-1;][]{2014AstL...40...65S} and searching for radioactive $^{56}Co$ decay lines after the explosion of the type Ia supernova SN 2014J in the galaxy M82 \citep{2014Natur.512..406C}. Based on these data, \cite{2016MNRAS.459..140M} constructed a deep survey of the M81 ﬁeld. Subsequently, having added the data of later observations, \cite{2023AstL...49....1M} obtained upper limits on the bolometric luminosity ($L_{\rm bol}\lesssim 10^{41}$~\lum) for the nuclei of 72 nearby dwarf galaxies located in the M81 field.

Regular \integral\ observations of the M81 field have been carried out at the requests of our group since 2019. The main objective of this deep survey is to search for strongly absorbed AGNs based on the set of \integral\ observational data in hard X-rays and the all-sky survey that has been conducted by the SRG observatory \citep{2021A&A...656A.132S} in softer X-rays since December 2019. Owing to the so long observations within this and previous programs (the total exposure time corrected for the dead time
is 19.2~Ms), by now a limiting sensitivity better than 0.2~mCrab\footnote{A flux of 1~mCrab in the 25$-$60~keV energy band corresponds to $9.73\times 10^{-12}$~\flux\ assuming the spectral model 10(E/1 keV)$^{-2.1}$ phot. cm$^{-2}$ s$^{-1}$ keV$^{-1}$.} has been achieved in the M81 field with the \ibis/ISGRI instrument \citep{2003A&A...411L.141L}. This is the deepest survey in the \integral\ extragalactic sky.

\section{DATA ANALYSIS}
\label{sec:data}

To construct a sky map, we used all of the available data obtained from March 2003 to August 2023 with the \ibis/ISGRI instrument within $18^{\circ}$ of the galaxy M81. We used the software package developed at the Space Research Institute of the Russian Academy of Sciences \citep{2010A&A...519A.107K,2014Natur.512..406C}.

\begin{figure}[t]
\centering
   \includegraphics[width=0.5\textwidth]{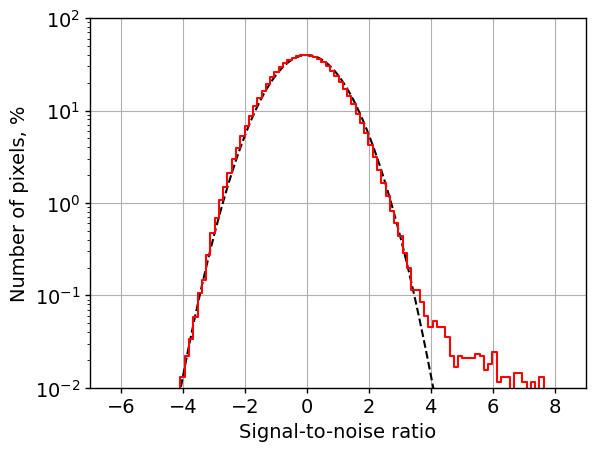}
     \caption{The distribution of pixels for the M81 ﬁeld (Fig.~\ref{fig:map}) in units of the signal-to-noise ratio (red histogram). The dashed line indicates the normal distribution with a standard deviation equal to unity.}
     \label{fig:snr}
\end{figure}

Because of the gradual degradation of the \ibis\ detectors with time and the related rise of the lower boundary of the effective sensitivity range from 17 to 25 keV, we chose the 25$-$60~keV energy band to construct the survey. This allows the influence of the systematic instrumental noise to be virtually eliminated but is accompanied by some loss of sensitivity \citep[for more details, see][]{2022MNRAS.510.4796K}.

Next, to produce the catalog of sources, we additionally bounded the survey region by the contour within which a sensitivity better than 0.72 mCrab ($6.8\times10^{-12}$~\flux) in the 25$-$60~keV energy band at a signiﬁcance level of $4\sigma$ is achieved. The central part of the field is speciﬁed by the coordinates RA, Dec = 146.3$^{\circ}$, 69.9$^{\circ}$ (FK5, epoch J2000), where a limiting sensitivity of 0.16~mCrab ($1.5\times10^{-12}$~\flux) is achieved. For a detection threshold of $4\sigma$ the expected number of false sources is much less than one object in the entire ﬁeld. The geometric area of the survey is 1004 deg$^{2}$, which roughly corresponds to a circle with a radius of 18$^{\circ}$.

As has been said above, owing to the rise of the lower boundary of the energy range, the sensitivity of the survey is not subjected to the systematic noise and is determined by the photon statistics. To prove this assertion, we constructed the distribution of all pixels in units of the signal-to-noise ratio shown in Fig.~\ref{fig:snr}. The properties of the distribution are well described by the Gaussian statistics, except for the wing in the positive region, where the contribution from X-ray sources is observed.

\begin{figure*}
\centering
   \includegraphics[width=0.99\textwidth]{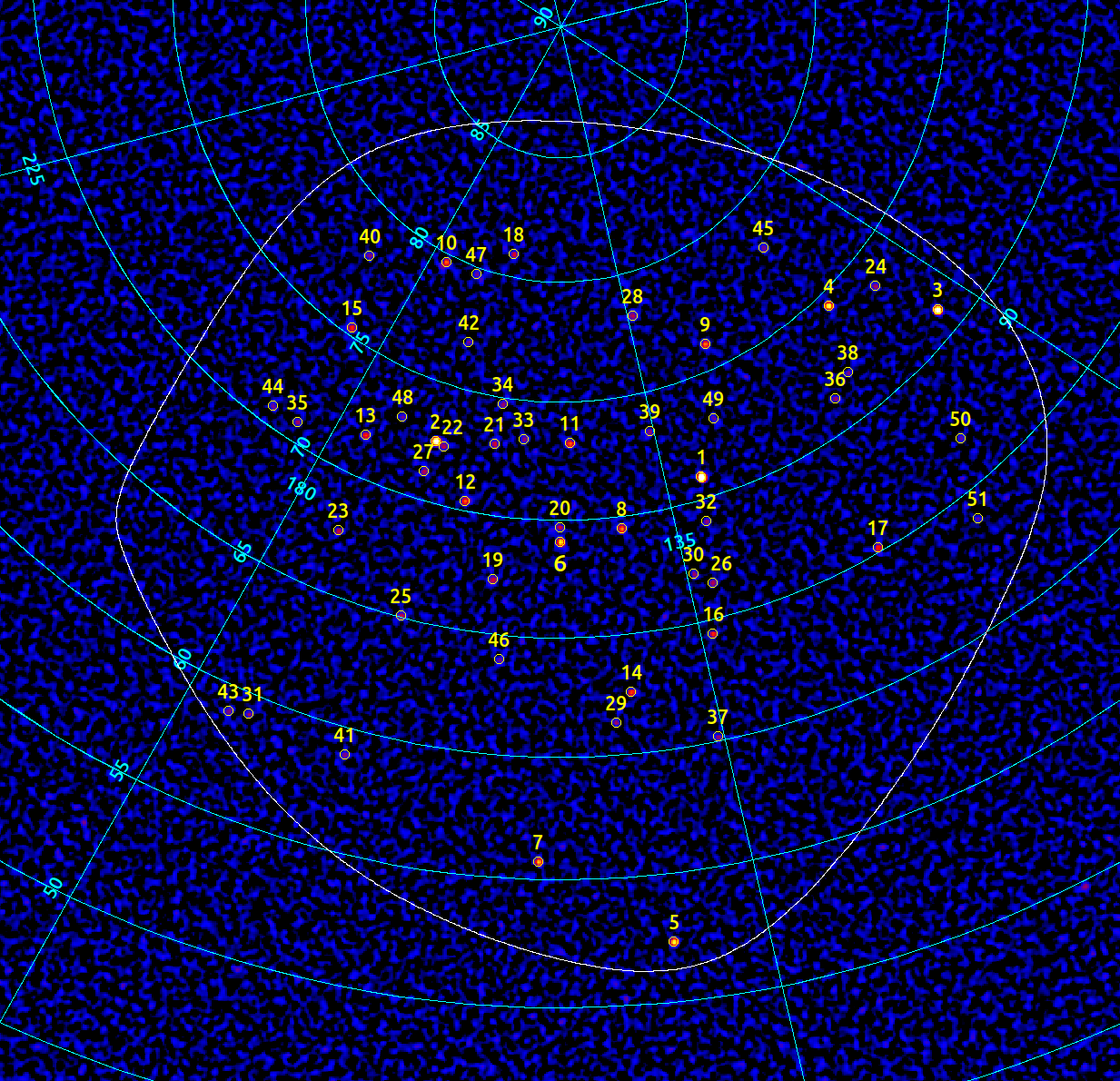}
     \caption{Map of the sky region around the galaxy M81 (source no. 6) in the 25$-$60~keV energy band from the \integral/\ibis\ data accumulated in an exposure time of 19.2~Ms (corrected for the dead time). The source numbers sorted by detection significance correspond to the numbers in Table~\ref{tab:catalog}. The white contour bounds the survey region in which the sensitivity exceeds 0.72~mCrab. The sources were detected in this region. The total image size is $48^{\circ}\times 48^{\circ}$. The coordinate grid maps the equatorial coordinate system; the north pole is seen in the upper part of the image.}
     \label{fig:map}
\end{figure*}

\begin{figure*}
\centering
   \includegraphics[width=0.99\textwidth]{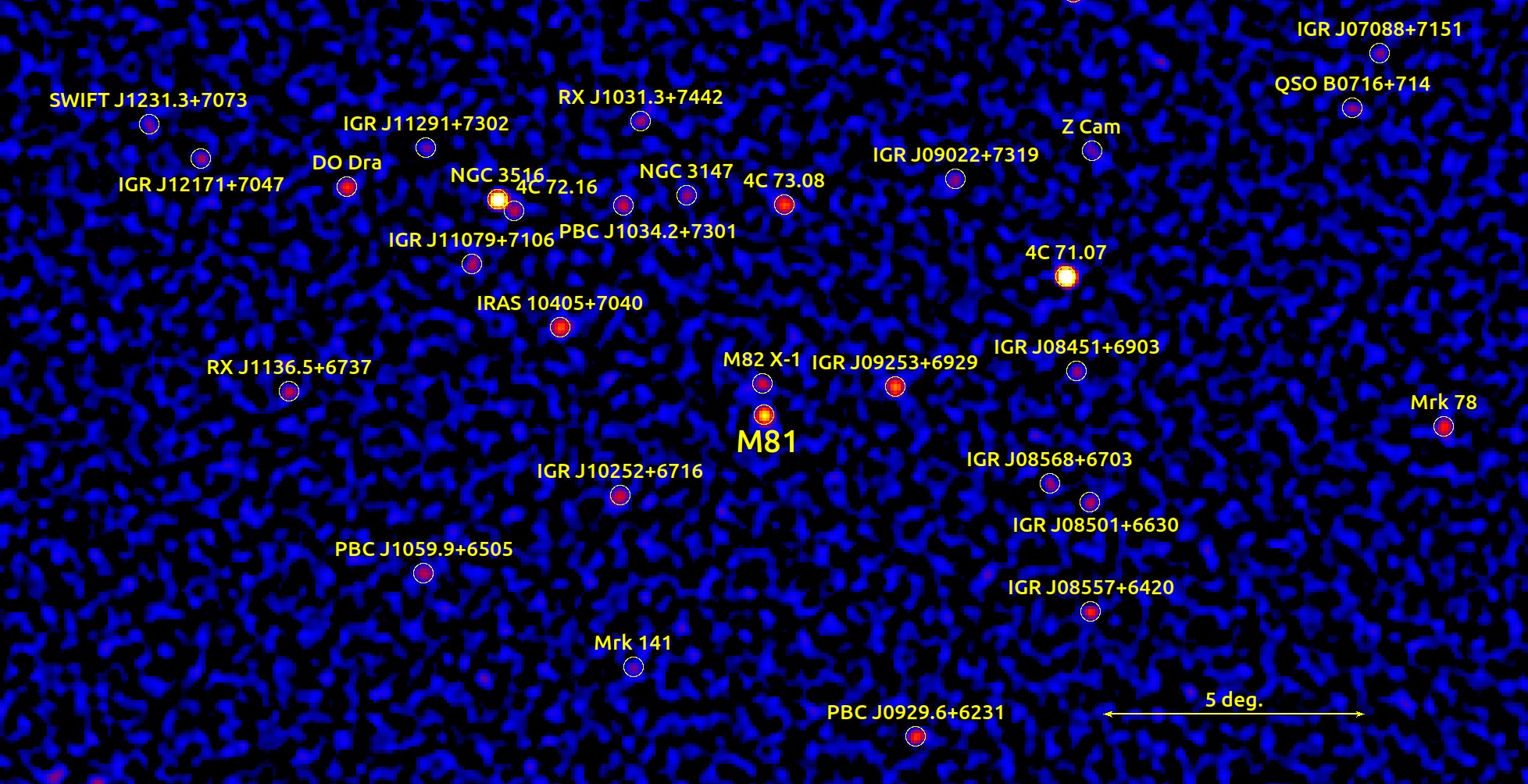}
     \caption{The central part of the zoomed X-ray image of the M81 ﬁeld with the source names.}
     \label{fig:zoom}
\end{figure*}

Figure~\ref{fig:map} shows the X-ray image of the M81 ﬁeld in the 25$-$60~keV energy band accumulated in the total exposure time of 19.2~Ms. Figure~\ref{fig:zoom} shows a zoomed image of the central part of the survey.

\section{THE CATALOG OF SOURCES}

The sources were detected using the procedures described in \cite{2022MNRAS.510.4796K}. Table~\ref{tab:catalog} presents the list of 51 objects detected at a signiﬁcance level higher than $4\sigma$. The positional accuracy of the sources detected with the \ibis\ coded-aperture telescope depends on their detection signiﬁcance \citep{2003A&A...411L.179G}. According to the estimates made in \cite{krivonos07allsky}, the 68\% conﬁdence intervals for a detection significance of 5$-$6, 10, and $>20\sigma$ are $2.1'$, $1.5'$, and $<0.8'$, respectively.

In our previous paper \citep{2016MNRAS.459..140M} we detected 37 sources in the M81 field at a significance level higher than $4.5\sigma$ based on \integral\ data. Thus, owing to the virtually doubled exposure (19.2 versus 9.7 Ms) and the slightly lowered detection threshold ($4\sigma$), we managed to detect much more objects in the same ﬁeld. Three sources, IGR~J11030+7027, IGR~J07563+5919, and IGR~J10380+8435, from the catalog of \cite{2016MNRAS.459..140M} were not detected in the current survey. This may be due to their variability. However, the possibility of a false detection of any of them must not be ruled out either. On the other hand, compared to \citep{2016MNRAS.459..140M}, fifteen new sources were added in the current survey.

We identified and classified the sources from the new catalog of the M81 field using publicly accessible astronomical databases, such as the SIMBAD Astronomical Database \citep{simbad} and the NASA/IPAC Extragalactic Database\footnote{Funded by NASA and maintained by the California Institute of Technology.} (NED), and astrophysical literature. For nearby extragalactic objects we used the distance estimates from the Extragalactic Distance Database \citep{tully23}. Our sample consists mostly of AGNs. These include 32 Seyfert galaxies, one LINER galaxy (i.e., weak activity is observed in the nucleus) -- M81, six blazars, and eight AGN candidates that will be discussed in detail below. In addition, the catalog contains one ultraluminous X-ray source (M82 X-1) and three cataclysmic variables in our Galaxy.

\begin{figure}[ht]
\centering
   \includegraphics[width=0.9\columnwidth]{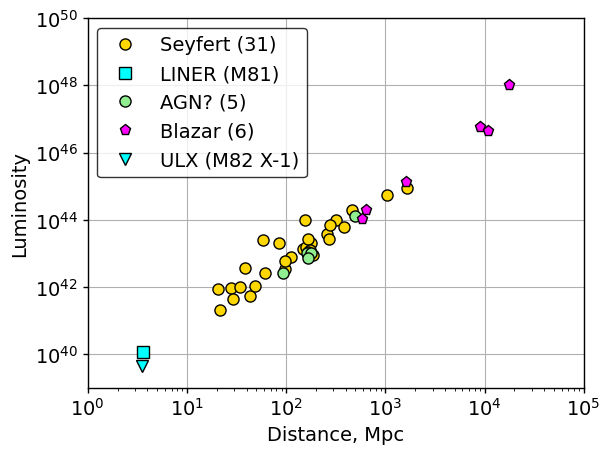}
     \caption{The distance -- luminosity (in the 25$-$60~keV energy band) diagram for the extragalactic objects in the M81 ﬁeld. The objects are divided into five categories: (1) Seyfert galaxies (31 objects), (2) LINER galaxies (one object M81), (3) blazars (six objects), (4) AGN candidates (AGN?, ﬁve objects with known redshifts), and (5) ultraluminous X-ray sources (ULX, one object M82 X-1).}
     \label{fig:lum}
\end{figure}

\begin{figure}[ht]
\centering
   \includegraphics[width=0.9\columnwidth]{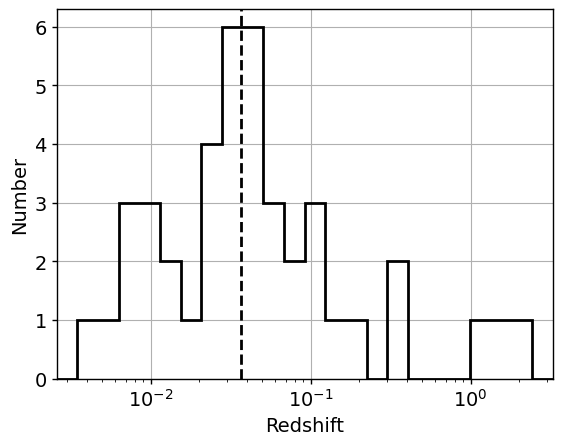}
     \caption{The redshift distribution of the Seyfert galaxies, AGN candidates, and blazars. The vertical dashed line indicates the median $z=0.0366$ for the sample of Seyfert galaxies and AGN candidates.}
     \label{fig:z}
\end{figure}

Figure~\ref{fig:lum} shows the distance -- luminosity (in the 25$-$60~keV energy band) diagram for the extragalactic sources in the \integral\ survey of the M81 field. Figure~\ref{fig:z} shows the redshift distribution of the AGNs and AGN candidates. When constructing these plots, we excluded the source IGR~J11058+5852 that apparently consists of two unresolved sources, SWIFT~J1105.7+5854A and SWIFT J1105.7+5854B -- known Seyfert galaxies located at diﬀerent redshifts (see below).

The median redshift of the conﬁrmed Seyfert galaxies and AGN candidates in the new catalog is $z_{\rm med}=0.0366$. Owing to the emission collimated toward us, the blazars, despite being rare in the Universe, can be detected from much greater distances and have much higher observed luminosities. For comparison, in the catalog of the 70-month Swift/BAT all-sky hard X-ray survey \citep{2017ApJS..233...17R} the median redshift of 731 AGNs, except for the blazars, is $z_{\rm med}=0.0367$, while in the catalog of the 17-year INTEGRAL all-sky survey \citep{2022MNRAS.510.4796K} $z_{\rm med} = 0.029$ based on a sample of 331 AGNs, except for the blazars. All of the mentioned surveys, including the deep \integral\ survey of the M81 ﬁeld, eﬀectively scan only the relatively nearby Universe (at distances less than several hundred Mpc). Therefore, the resulting AGN statistics is determined by a combination of the exposure map and the large-scale structure of the Universe in the region of a specific survey.

\section{INTERSECTIONS WITH OTHER CATALOGS}

Most of the objects in the new catalog of the \integral\ survey of the M81 ﬁeld have already been known previously as X-ray sources. We matched the catalog with the catalogs of a number of all-sky X-ray surveys, searching for counterparts within the error circle around the sources from our catalog.

An up-to-date (based on 17 years of observations) catalog of sources of the \integral\ all-sky survey detected in the 17$-$60~keV energy band was presented in \citep{2022MNRAS.510.4796K}. Thirty three sources are common to the two catalogs. Three sources from the catalog of \cite{2022MNRAS.510.4796K}, IGR~J10380+8435, IGR~J11030+7027, and IGR~J07563+5919, did not exceed the detection threshold in the current M81 survey, with these sources having been detected previously in this field in \cite{2016MNRAS.459..140M}, where the 17$-$60~keV energy band was used. These sources probably have relatively soft X-ray spectra, which did not allow them to be detected at a sufficiently high significance level after the rise of the lower boundary of the energy band from 17 to 25 keV in the current survey.

Among the X-ray surveys of other observatories, the all-sky survey with the BAT telescope \citep{bat} onboard the Gehrels Swift observatory \citep{swift} already mentioned above is most similar in energy band. A cross-correlation with the catalog of sources detected over 105 months in the energy range 14$-$195~keV \citep{Oh18} revealed 38 intersections with the new catalog of the \integral\ M81 survey. The remaining 13 sources from our catalog are absent in the catalog of \cite{Oh18}.

Recently, a new catalog of sources detected with the Mikhail Pavlinsky ART-XC telescope onboard the Spectrum–RG observatory in the 4$-$12~keV energy band based on the sum of five all-sky surveys has been released (Sazonov et al. 2024). We found 19 intersections with this catalog in the current \integral\ catalog of the M81 ﬁeld.

Six sources detected in the \integral\ survey of the M81 field have not been detected previously in any of the all-sky X-ray surveys mentioned above.

\section{NEW SOURCES DETECTED DURING THE SURVEY}

A number of sources in the new catalog (see Table~\ref{tab:catalog}) have previously been unknown or poorly studied. We attempted to identify the pre-classify these objects based on the information available in various astronomical catalogs. These cases are discussed below.

\subsection{No. 26. IGR J08501+6630}

In the position error circle of this source the \ibis\ instrument distinguishes the edge-on galaxy MCG+11-11-029 at redshift $z = 0.0370$. The radio source ILT~J085033.04+662916.2 from the LOFAR survey \citep{lofar} can be associated with it. The infrared (IR) color of the galaxy from the ALLWISE catalog \citep{allwise}, $W1-W2\approx0.2$, is more typical for ordinary galaxies than for AGNs. Nevertheless, it can be assumed that we are dealing with an absorbed AGN, since its nucleus may be hidden from the observer behind a thick layer of interstellar matter due to the orientation of the galaxy in the sky.

\subsection{No. 30. IGR J08568+6703}

This source was included in the unpublished third Palermo-BAT Catalog, 3PBC\footnote{\url{http://bat.ifc.inaf.it/bat_catalog_web/66m_bat_catalog.html}} \citep{2010A&A...524A..64C}. Based on Swift/XRT observations, it was identiﬁed with the soft X-ray source 2SXPS~J085656.3+670255 \citep{2sxps} that, in turn, can be identiﬁed with the extended object WISEA~J085656.49+670257.3 detected at wavelengths from the ultraviolet \citep[GALEX,][]{galex} to radio \citep[LOFAR, VLASS,][]{lofar,vlass}. Its IR color $W1-W2\approx0.9$ clearly points to the presence of an active nucleus.

\subsection{No. 32. IGR J08451+6903}

In the \ibis\ position error circle there is the galaxy EDA~24565 at $z=0.0405$ with an IR color $W1-W2\approx 0.5$ typical for AGNs, which is also a radio source \citep[NVSS,][and other surveys]{nvss}. Thus, this is yet another AGN candidate.

\subsection{No. 38. IGR J07088+7151}

At this place there is the galaxy cluster Abell~565 ($z=0.1054$) that manifests itself as the extended soft X-ray source RXC~J0708.1+7151 \citep{noras} = XMMSL2~J070808.9+715150 \citep{saxton08}. The soft X-ray flux is comparable to the hard X-ray flux recorded by \integral. However, the cluster gas is not hot enough \citep[$kT = 3.68 \pm
0.12$~keV,][]{xu22} for the hard X-ray emission to be associated with the thermal gas radiation. We can cautiously assume that IGR~J07088+7151 is an AGN located in the cluster Abell~565.

\subsection{No. 39. IGR J09022+7319}

The galaxy MCG+12-09-029 ($z=0.0371$) is distinguished in the \ibis\ position error circle. It is not remarkable from the standpoint of the IR color ($W1-W2=0.0$) and any other photometric signatures of nuclear activity. Nevertheless, in the absence of better hypotheses we assign this source to the AGN candidates.

\subsection{No. 41. IGR J11058+5852}

This source is probably a superposition of two closely spaced (in the sky) hard X-ray sources, SWIFT~J1105.7+5854A and SWIFT~J1105.7+5854B, from the catalog of the Swift/BAT all-sky survey \citep{Oh18}. Both these objects are Seyfert galaxies: Z~291-28 ($z=0.0476$) and 2MASS~J11053761+5851208 ($z=0.1910$), respectively. Based on \ibis\ data, we failed to separate this pair of sources.

\subsection{No. 45. IGR J06571+7802}

In the \ibis\ position error circle there is the radio source NVSS~J065745+780059 detected in many surveys. It is identiﬁed with the faint IR source WISEA~J065745.76+780101.2 ($W1-W2 \sim 0.5$). Thus, this is yet another AGN candidate.

\subsection{No. 48. IGR J11291+7302}

The galaxy UGC 6473 ($z=0.0214$) is distinguished in the \ibis\ position error circle. Radio emission (NVSS~J112857+730202) is detected from it. Its IR color $W1-W2 \approx 0.1$ corresponds to ordinary galaxies but does not rule out the presence of an active nucleus, given the relative proximity and low luminosity of the object. Therefore, we assign this object to the AGN candidates.

\subsection{No. 50. IGR J06507+6647}

In the \ibis\ position error circle there is the bright radio source NVSS~J065026+664937. The IR source WISEA~J065026.72+664942.3 with $W1-W2 \approx 0.7$ can be associated with it. These signatures point to the presence of an AGN.

\subsection{No. 51. IGR J07064+6353}

This source is identified with the source SRGA~J070637.0+635109 from the catalog of the
SRG/ART-XC all-sky survey \citep[][Sazonov et al. 2024]{pavlinskiy2022}. Recently, its optical spectrum has been obtained, which has allowed one to classify it as a Seyfert 1.8 galaxy and to measure its redshift: $z=0.0140$ \citep{2022AstL...48...87U}.

\section{CONCLUSIONS}
\label{sec:summary}

Based on long-term \integral\ observations, we carried out a deep X-ray survey of the M81 field. A record sensitivity of 0.16~mCrab in the 25$-$60~keV energy band at a detection significance of $4\sigma$ was achieved in the central part of the field owing to the accumulated exposure of 19.2~Ms. The total area of the survey is 1004 deg$^2$ at a sensitivity level better than 0.72~mCrab ($4\sigma$).

The catalog of X-ray sources detected at a significance level higher than $4\sigma$ contains 51 objects most of which are AGNs. Six sources have not been detected previously in any of the X-ray surveys. According to the available data, all of them and two more sources that have already been entered previously into the \integral\ survey catalogs can be AGNs, including those with strong internal absorption. Pointed X-ray observations and optical spectroscopy are required to test this assumption.

\section{ACKNOWLEDGMENTS}

This study is based on observations within the \integral\ project of the European Space Agency being implemented in cooperation with Russia and USA. The data were retrieved from the European and Russian \integral\ Science Data Centers. We thank E.M. Churazov who developed the IBIS/INTEGRAL data analysis methods and are grateful to the Max Planck Institute for Astrophysics (Germany) for the computational support.

\begin{landscape}
\begin{table}[ht]
\small
\centering
\caption{The catalog of X-ray sources detected in the M81 ﬁeld based on \integral/\ibis\ data in the 25$-$60 keV energy band. The sources are listed in order of decreasing detection signiﬁcance ($S/N$).}
\begin{tabular}{l|l|c|c|c|c|c|c|l|l} 
\hline
 No. & Name & RA & Dec & Flux & $S/N$ & $z$ & Type & Surveys$^{\rm a)}$ & Notes, references \\
     &          & (J2000) & (J2000) & nCrab & &  &  & & \\
\hline
1 & 4C 71.07 & 130.3376 & 70.9024 & 2.95 $\pm$ 0.05 & 63.7 & 2.1720 & Blazar & 1-4 & \\ 
2 & NGC 3516 & 166.7135 & 72.5649 & 2.13 $\pm$ 0.05 & 44.3 & 0.0088 & Sy1.5 & 1-4 & \\
3 & Mrk 3 & 93.8886 & 71.0396 & 6.07 $\pm$ 0.15 & 39.3 & 0.0135 & Sy2 & 1-4 & \\
4 & Mrk 6 & 103.0381 & 74.4317 & 2.35 $\pm$ 0.10 & 23.3 & 0.0195 & Sy2 & 1-4 & \\
5 & Mrk 110 & 141.3090 & 52.2853 & 3.51 $\pm$ 0.16 & 22.2 & 0.0350 & Sy1 & 1-4 & \\
6 & M81 & 148.8602 & 69.0720 & 0.79 $\pm$ 0.04 & 19.9 & & LINER & 1-4 & D=3.61 Mpc\\
7 & NGC 3079 & 150.4961 & 55.6897 & 1.82 $\pm$ 0.10 & 17.4 & 0.0037 & Sy2 & 1-3 & D=20.61 Mpc \\
8 & IGR J09253+6929 & 141.4370 & 69.4772 & 0.55 $\pm$ 0.04 & 13.7 & 0.0398 & Sy1.5 & 1-4 & \\ 
9 & PG 0804+761 & 122.8281 & 76.0597 & 0.78 $\pm$ 0.06 & 12.9 & 0.1000 & Sy1 & 1-4 & \\
10 & 1ES 1141+79.9 & 176.3406 & 79.6691 & 1.02 $\pm$ 0.08 & 12.4 & 0.0065 & Sy1.2 & 1-4 & \\
11 & 4C 73.08 & 147.4671 & 73.2467 & 0.49 $\pm$ 0.04 & 11.8 & 0.0580 & Sy2 (NLRG) & 1-3 & \\
12 & IRAS 10405+7040 & 161.0241 & 70.4123 & 0.51 $\pm$ 0.04 & 11.7 & 0.0336 & Sy2 & 1-3 & \\
13 & DO Dra & 175.9362 & 71.7171 & 0.63 $\pm$ 0.06 & 10.6 & & CV & 1-4 & \\
14 & PBC J0929.6+6231 & 142.3670 & 62.5556 & 0.54 $\pm$ 0.05 & 10.5 &0.0256 & Sy2 & 1-3 & \\ 
15 & Mrk 205 & 185.4558 & 75.3079 & 0.84 $\pm$ 0.08 & 10.1 & 0.0709 & Sy1 & 1-4 & \\
16 & IGR J08557+6420 & 133.8026 & 64.3936 & 0.51 $\pm$ 0.05 & 9.9 & 0.0362 & Sy2 & 1-3 & \\
17 & Mrk 78 & 115.7189 & 65.1813 & 0.80 $\pm$ 0.09 & 9.2 & 0.0379 & Sy2 & 3 & \\
18 & 6C 103912+811006 & 161.2189 & 80.9152 & 0.64 $\pm$ 0.08 & 8.2 & 1.2600 & Blazar & 1-3 & \\
19 & IGR J10252+6716 & 156.3028 & 67.2989 & 0.32 $\pm$ 0.04 & 7.7 & 0.0386 & Sy2 & 1-3 & \\
20 & M82 X-1 & 148.9556 & 69.6963 & 0.30 $\pm$ 0.04 & 7.6 & & ULX & 1-4 & D=3.53 Mpc \\
21 & PBC J1034.2+7301 & 158.4465 & 73.0057 & 0.32 $\pm$ 0.04 & 7.3 & 0.0220	& Sy2 & 1-3 & \\
22 & 4C 72.16 & 165.4542 & 72.4318 & 0.34 $\pm$ 0.05 & 7.3 & 1.4600 & Blazar &	1,2 & \\
23 & RX J1136.5+6737 & 174.1284 & 67.5920 & 0.43 $\pm$ 0.06 & 7.2 & 0.1342 & Blazar & 1-4 & \\ 
24 & MU Cam & 96.4044 & 73.5744 & 0.93 $\pm$ 0.13 & 7.2 & & CV & 1-4 & \\ 
25 & PBC J1059.9+6505 & 165.0025 & 65.0632 & 0.35 $\pm$ 0.05 & 6.7 & 0.0836 & Sy2 & 3,4 & \\
26 & IGR J08501+6630 & 132.5434 & 66.4836 & 0.32 $\pm$ 0.05 & 6.6 & 0.0370 & AGN? & 1,2 & MCG+11-11-029 \\
27 & IGR J11079+7106 & 167.0707 & 71.1865 & 0.32 $\pm$ 0.05 & 6.6 & 0.0600 & AGN & 1-3 & \\ 
28 & NGC 2655 & 133.8822 & 78.2078 & 0.39 $\pm$ 0.06 & 6.6 & 0.0050 & Sy2 & 1-3 & \\
29 & IRAS 09320+6134 & 143.9668 & 61.3389 & 0.35 $\pm$ 0.06 & 6.2 & 0.0394 & Sy1 & 1-3 & \\
30 & IGR J08568+6703 & 134.2313 & 67.0334 & 0.28 $\pm$ 0.05 & 6.1 & & AGN? & & PBC J0857.3+6704 \\ 
31 & SBS 1136+594 & 174.7757 & 59.1829 & 0.79 $\pm$ 0.13 & 5.9 & 0.0612 & Sy1 & 3,4 & \\
32 & IGR J08451+6903 & 131.3212 & 69.0492 & 0.27 $\pm$ 0.05 & 5.9 & 0.0405 & AGN? & & LEDA 24565 \\
33 & NGC 3147 & 154.2565 & 73.3686 & 0.25 $\pm$ 0.04 & 5.8 & 0.0099 & Sy2& 3 & \\
34 & RX J1031.3+7442 & 158.1377 & 74.7052 & 0.27 $\pm$ 0.05 & 5.8 & 0.1230 & Blazar & 3,4 & \\
35 & IGR J12171+7047 & 184.3606 & 70.7967 & 0.46 $\pm$ 0.08 & 5.7 & 0.0067 & AGN & 1,2 & \cite{malizia20} \\ 
\hline
\end{tabular}
\label{tab:catalog}
\begin{flushleft}
    $^{\rm a)}$The source is also present in the following catalogs: 1 -- the \integral\ M81 ﬁeld survey \citep{2016MNRAS.459..140M}, 2 -- the 17-year INTEGRAL all-sky survey \citep{2022MNRAS.510.4796K}, 3 -- the 105-month Swift/BAT all-sky survey \citep{Oh18}, 4 -- the SRG/ART-XC all-sky survey (Sazonov et al. 2024).
\end{flushleft}
\end{table}
\end{landscape}

\begin{landscape}
\setcounter{table}{0}
\begin{table}[ht]
\small
\centering
\caption{(Contd.)}
\begin{tabular}{llccccccll} 
\hline
 No. & Name & RA & Dec & Flux & S/N & $z$ & Type & Surveys$^{\rm a)}$ & Notes \\
     &          & (J2000) & (J2000) & mCrab & &  &  & & \\
\hline
36 & QSO B0716+714 & 110.4952 & 71.3127 & 0.44 $\pm$ 0.08 & 5.5 & 0.3100 & Blazar & 1,2 &\\
37 & Mrk 18 & 135.4907 & 60.1509 & 0.39 $\pm$ 0.07 & 5.5 & 0.0111 & Sy2 & 3 & \\
38 & IGR J07088+7151 & 107.1802 & 71.8305 & 0.48 $\pm$ 0.09 & 5.5 & 0.1054 & AGN? & & In cluster Abell 565? \\
39 & IGR J09022+7319 & 135.5513 & 73.3271 & 0.24 $\pm$ 0.04 & 5.4 & 0.0371 & AGN? & & MCG+12-09-029 \\
40 & IGR J12418+7805 & 190.5876 & 78.1117 & 0.56 $\pm$ 0.10 & 5.3 & 0.0221 & Sy1.9 & 1,2 & \\ 
41 & IGR J11058+5852 & 166.4717 & 58.8819 & 0.50 $\pm$ 0.10 & 5.2 & 0.1910 & Sy1 & 1-3 & 1) SWIFT J1105.7+5854B \\ 
& & & & & & 0.0476 & Sy2 & & 2) SWIFT J1105.7+5854A \\ 
42 & PG 1100+772 & 166.1402 & 76.9426 & 0.28 $\pm$ 0.06 & 4.9 & 0.3115 & Sy1 & 3,4 & \\
43 & SWIFT J1145.2+5905 & 176.3217 & 58.9800 & 0.74 $\pm$ 0.15 & 4.9 & 0.0079 & Sy2 & 3 & \\
44 & SWIFT J1231.3+7073 & 187.9397 & 70.7567 & 0.45 $\pm$ 0.09 & 4.8 & 0.2080 & Sy1.2 & 1-3 & \\ 
45 & IGR J06571+7802 & 104.3319 & 77.9851 & 0.48 $\pm$ 0.10 & 4.6 & & AGN? & 1,2 & NVSS J065745+780059 \\
46 & Mrk 141 & 154.7628 & 63.9507 & 0.22 $\pm$ 0.05 & 4.6 & 0.0417 & Sy1 & 3 & \\
47 & PBC J1113.6+7942 & 168.8534 & 79.7113 & 0.34 $\pm$ 0.07 & 4.6 & 0.0372 & Sy2 & 1,3 &  \\
48 & IGR J11291+7302 & 172.2946 & 73.0448 & 0.25 $\pm$ 0.05 & 4.5 & 0.0214 & AGN? & & UGC 6473 \\ 
49 & Z Cam & 126.1507 & 73.0603 & 0.22 $\pm$ 0.05 & 4.3 & & CV & 3 & \\
50 & IGR J06507+6647 & 102.6886 & 66.7964 & 0.56 $\pm$ 0.13 & 4.2 & & AGN? & & NVSS J065026+664937\\ 
51 & IGR J07064+6353 & 106.6215 & 63.8969 & 0.60 $\pm$ 0.15 & 4.1 &  0.0140 & Sy1.8 & 4 & SRGA J070637.0+635109\\ 
\hline
\end{tabular}

\begin{flushleft}
    $^{\rm a)}$The source is also present in the following catalogs: 1 -- the \integral\ M81 ﬁeld survey \citep{2016MNRAS.459..140M}, 2 -- the 17-year INTEGRAL all-sky survey \citep{2022MNRAS.510.4796K}, 3 -- the 105-month Swift/BAT all-sky survey \citep{Oh18}, 4 -- the SRG/ART-XC all-sky survey (Sazonov et al. 2024).
\end{flushleft}
\end{table}
\end{landscape}

\section{FUNDING}

This work was supported by RSF grant no. 19-12-00396.

\section{CONFLICT OF INTEREST}

The authors of this work declare that they have no conflicts of interest.


\bibliographystyle{astl}
\bibliography{biblio} 


{\it Translated by V. Astakhov}

{\it Latex style was created by R. Burenin}

\label{lastpage}

\end{document}